\documentclass[prl,twocolumn]{revtex4}

\usepackage{graphicx}
\usepackage{dcolumn}
\usepackage{bm}
\def\ä{\"{a}}
\def\ü{\"{u}}
\def\ö{\"{o}}
\def\Ä{\"{A}}
\def\Ü{\"{U}}
\def\Ö{\"{O}}

\begin{document}

\preprint{APS/123-QED}

\title{Searching atomic spin contrast on nickel oxide (001) by force microscopy}

\author{M. Schmid}
\author{F. J. Giessibl}
\email{franz.giessibl@physik.uni-augsburg.de}
\author{J. Mannhart}


\affiliation{%
Universit\"at Augsburg, Institute of Physics, Electronic
Correlations and Magnetism, Experimentalphysik VI,
Universit\"atsstrasse 1, D-86135 Augsburg, Germany.
}%

\date{\today, submitted to Physical Review B}

\begin{abstract}
The (001) surface of NiO, an antiferromagnet at room temperature,
was investigated under ultra-high vacuum conditions with frequency
modulation atomic force microscopy (FM-AFM). The antiferromagnetic
coupling between ions leads to a spin superstructure on (001)
surfaces. Exchange interaction between the probe of a force
microscope and the NiO (001) surface should allow to image spin
superstructures in real space. The surface was imaged with three
different probing tips: nonmagnetic W tips, ferromagnetic Co tips
and antiferromagnetic NiO tips - and atomic resolution was
achieved with all three of them in various distance regimes and in
several channels. Evidence for spin contrast was obtained in
experiments that utilize NiO tips and oscillation amplitudes in
the \AA-regime, where optimal signal-to-noise ratio is expected.
The spin contrast is weaker than expected and only visible in
Fourier space images.
\end{abstract}

\maketitle

\section{Introduction}

The electronic and mechanical properties of matter are dominated
by the Coulomb interaction resulting from the charge of the
electrons. In contrast, the magnetic interaction of the spin of
the electrons plays a minor role. The dipole-dipole interaction of
single electronic spins for typical interatomic distances is only
on the order of a few $\mu$eV and electrostatic energies between
two electrons are $10^6$-times larger. While the direct
interaction energy between spins is small, the Pauli principle
constrains the symmetry of wave functions of two-electron states
depending on spin: the spatial part of a spin-singlet state must
keep its sign with particle exchange, while a spin-triplet state
flips the sign of the spatial part of the wave function with
particle exchange. In H$_2$, the energetic difference between its
two electrons occupying singlet- vs. triplet states (exchange
interaction) amounts to several eV's \cite{Baym}. Therefore, spin
is important in solids, and it is important to establish tools
that allow to analyze spin orientation on surfaces. For conductive
samples, spin-polarized scanning tunneling microscopy
\cite{Bode_2003} is a powerful tool to image the spin orientation
of surface atoms within magnetic domains or even antiferromagnetic
surfaces with atomic resolution \cite{Heinze_2000}. Recently, the
spin of a single magnetic ion has been measured by scanning
tunneling spectroscopy \cite{Heinrich_2004}. However, the spin
orientation is also of interest in insulating materials such as
magnetic oxides. Insulators can be imaged by atomic force
microscopy (AFM) \cite{binnig_86}, and magnetic force microscopy
(MFM), a variation of AFM, allows magnetic imaging through the
magnetic dipole interaction of magnetic domains in the probe tip
and in the sample. Because of the weak dipole-dipole interaction,
many spins comprising larger domains are necessary to measure
magnetic dipole forces and the spatial resolution of magnetic
force microscopy is limited to some 10\,nm. In contrast, exchange
interaction can lead to spin-dependent interaction energies of up
to 100\,meV, and atomic imaging of exchange interactions on {\it
ferromagnetic} samples by AFM has been proposed early after atomic
resolution AFM became available \cite{Nakamura_1999}. For two
reasons, antiferromagnetic samples are attractive to probe the
possibility of exchange force measurements: a) they provide
well-defined magnetic contrast over small lateral distances and b)
the disturbing magnetic dipole interaction between a magnetic tip
and an antiferromagnetic sample is weak and decays exponentially
with distance. NiO (001) is a good choice for a test sample,
because it is antiferromagnetic at room temperature and (001)
surfaces with excellent flatness and cleanliness can be prepared
relatively easily by cleavage in ultrahigh vacuum. Because of its
magnetic properties, NiO is used as a pinning layer in spin valves
and has been instrumental in the study of metal-insulator
transitions \cite{Metal_Insulator_Transitions_1998}. Several
groups have studied NiO (001) by atomic force microscopy and
obtained atomic images of the surface \cite{NiO_Hosoi_2000,
NiO_Hosoi_2001, NiO_Hosoi_2004, NiO_Wiesendanger_2001} and
performed spectroscopy \cite{NiO_Wiesendanger_2002,
NiO_Wiesendanger_2003, NiO_Hoffmann}, but a clear-cut proof of the
expected spin contrast is lacking. It has been proposed that spin
contrast can only be observed at very small tip-sample distances
\cite{NiO_expectations_Foster}, but this distance regime is
difficult to reach with conventional AFM with soft cantilevers
(spring constant $k\approx$ 40\,N/m) and large amplitudes
($A\approx$ 10\,nm). Also, conventional cantilevers are only
available made from Si and magnetic layers need to be deposited on
the cantilever which increases the tip radius. Here, we use stiff
cantilevers with $k\approx$ 4\,kN/m that can be operated at
extremely small distances, even in the repulsive regime.
Nevertheless, we did not see spin contrast even for very small
distances using ferromagnetic tips. We argue below that the
exchange interaction between ferromagnetic metal tips and NiO
might be much smaller than exchange interaction in NiO bulk. We
have therefore built force sensors that are equipped with NiO
crystal tips and use them to image NiO(001). We use analysis in
Fourier space to determine the extent of spin polarization and
find some evidence in selected experiments using NiO tips.

\section{Experimental}

Nickel oxide crystalizes in the rock salt structure with a lattice
constant of $a_0=4.17\,\rm\AA{}$. The spins are localized at the
Ni sites and are pointing to one of the six possible $\langle
1\overline21\rangle$ directions \cite{NiO_spindirection}. NiO is
an antiferromagnet with a N\'eel temperature well above room
temperature at $T_{\rm N}=525$\,K. Within the (111) planes, the
spins couple ferromagnetically, the coupling between neighboring
(111) planes is antiferromagnetic (see Fig.\,1). The intersection
of these planes with the (001) surface yields diagonals with
parallel spin alignment, where neighboring lines have opposite
spin directions.

\begin{figure}
\begin{center}
\includegraphics[clip=true, width=0.38\textwidth]{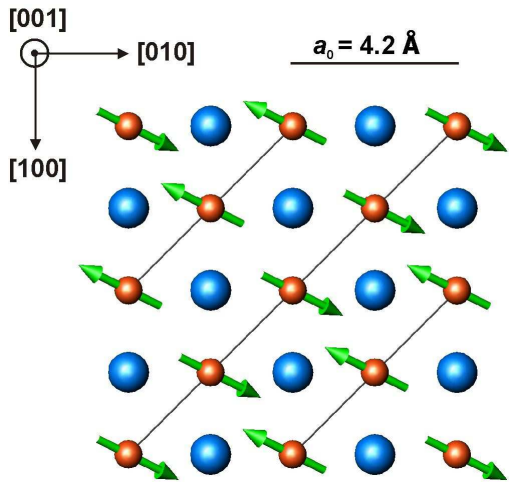}
\end{center}
\caption{(color online) NiO structure (spins located at the nickel
sites): the top view onto the (001) surface shows ferromagnetic
rows in [1$\overline 1$0] direction which couple
antiferromagnetically along the [110] direction.}
\end{figure}

The samples used in our experiments were single crystalline blocks
of NiO (SurfaceNet, Rheine, Germany). They were cut to bars of
about $2\times 4 \times 10$ mm$^3$ and mounted on a plate to allow
sample transfer from ambient conditions to vacuum and {\it in
situ} sample preparation. A gold layer of about 300\,nm thickness
was sputtered onto the samples to support the discharging of the
surface right after cleavage. To obtain flat and clean surfaces,
the crystals were cleaved {\it in-situ} with a UHV cleaving device
\cite{cleaver}. All experiments were performed at room temperature
at a pressure of $\approx 8\cdot 10^{-11}$\,mbar. Stable atomic
imaging could be achieved for up to four days from the time the
cleave was initiated, after that, contamination became visible
clearly.

Several estimations of the expected exchange interaction between a
magnetic tip and an antiferromagnetic sample surface have been
published. First-principle calculations for two magnetic Fe(100)
thin films with a distance in the range of the lattice constant
yield $E_{\rm ex}\approx 10\,\rm meV$ and $F_{\rm ex}\approx
0.1\,\rm nN$ \cite{NiO_expectations_Mukasa97}. A modelling of the
NiO(001) surface interacting with a spin-polarized H atom (weakly
reactive) and a spin-polarized Fe atom (strongly reactive) finds
that the difference in force over opposite spin atoms should be
detectable with the AFM for a tip-sample distance smaller than
4\,$\rm \AA{}$ or for imaging close to the repulsive regime
\cite{NiO_expectations_Foster}. However, at such short distances,
the chemical bonding forces can become strong and it was
speculated that ion instabilities may become apparent. Elongations
of the tip and the sample atomic bonds are no longer negligible
and atoms may even become displaced. They may lead to the loss of
atomic resolution before the marginal tip-sample distance for
detecting the exchange force is reached. Weakly reactive tips are
less affected by these instabilities. For bulk NiO, K\ödderitzsch
{\it et al.} \cite{NiO_exchange_interactions} have calculated that
the AF$_2$ antiferromagnetic structure displayed in Fig. 1 has a
bonding energy that is lower by 116\,meV per Ni-O atom pair than a
ferromagnetic spin arrangement. The Gibbs free energy of formation
for NiO is 211\,kJ/mol at room temperature \cite{CRC},
corresponding to 2.19\,eV per Ni-O atom pair. Because every ion in
NiO has six neighbors and every bond is shared by two ions, we
estimate a bonding energy of 730\,meV per bond. Therefore, the use
of a NiO tip to probe NiO(001) promises to provide large
spin-dependent contrast where the short-range bonding force varies
by 116\,meV/730\,meV = 16\%. The range of the exchange forces is
expected to be similar to the range of chemical bonds with
$\lambda_{\rm ex} \approx 0.1$\,nm.

Previous atomically resolved imaging experiments of NiO surfaces
all have parameters in the following ranges: oscillation
amplitudes of several nm and cantilevers with $k\approx 40$\,N/m
oscillating at frequencies of some hundreds of kHz
\cite{NiO_Hosoi_2000,NiO_Hosoi_2001,NiO_Hosoi_2004,
NiO_Wiesendanger_2001,NiO_Wiesendanger_2002,
NiO_Wiesendanger_2003,NiO_Hoffmann}. Optimal signal-to-noise ratio
is expected for oscillation amplitudes $A\approx \lambda$
\cite{SNR}, where $\lambda$ is the range of the interaction that
is to be probed. Because of stability requirements, $k\cdot A$ has
to exceed a critical value \cite{basic} and a large stiffness is
required for stable operation at small amplitudes. For this
purpose, the self-sensing quartz cantilever qPlus
\cite{giessibl_00}, which is based on a commercial tuning fork and
can be operated as is with oscillation amplitudes in the range of
several Angstroms, was modified for operation at even smaller
amplitudes. The stiffness of the prongs of the tuning fork is
given by $k = Ewt^3/4L^3$, where $L,t,w$ and $E$ are the length,
the thickness, the width and Youngs modulus of the prongs,
respectively. The modification involved a shortening of the prongs
by cutting them with a diamond wire saw, changing $k$ from
1800\,N/m to $\approx 4000\,{\rm N/m}$ and $f_0$ from $\approx
20\,{\rm kHz}$ to $\approx 40\,{\rm kHz}$. Stable oscillation at
amplitudes of $A\approx 1\,\rm \AA{}$ and below became possible
with these \lq extra stiff\rq{} qPlus sensors. Compared to the
cantilevers of conventional AFM, $k$ is increased by 100 allowing
a decrease of $A$ by a factor of 1/100. As a consequence,
additional to the advantage of attenuated long-range background
forces, qPlus extra stiff sensors promise to provide an increased
frequency shift and thus higher resolution on small scale.

The probe tips are important in AFM. Because of the large size and
rigidity of our qPlus force sensors, a wide variety of tips can be
mounted. Etched metal tips (e.g. W) as known from scanning
tunneling microscopy are standard, but cobalt was chosen as a
ferromagnetic tip material. Among the ferromagnetic elements it
shows the weakest reactivity which facilitates stable imaging
close to the sample surface. The etching was performed with a
$50\%$ solution of HNO$_3$. We also prepared antiferromagnetic
tips made from NiO for reasons that are outlined below. NiO tips
were prepared by cleaving larger crystals \textit{ex situ} and
searching for sharply pointed crystallites with sizes of roughly
$50 \mu$m$\times 50 \mu$m$\times 250 \mu$m. Annealing by electron
bombardment in UHV is difficult for an insulator like NiO.
Therefore, we attempted to clean the tips {\it in situ} by
scratching along the (NiO) surface.

In typical AFM images of ionic crystals, only one type of ion
appears as a protrusion, and the other type is imaged as a
depression. It depends on the tip whether Ni or O ions are imaged
as protrusions in AFM images of NiO (001). Momida et al.
\cite{Momida_2005} argue that oxygen atoms appear as bright
protrusions when using metal tips because metals react more
strongly with oxygen than other metals. However, this issue and
the identity of the tip atom and crystallographic environment
constitute uncertainties in the image interpretation which will
have to be discussed. Nevertheless, even if the oxygen atoms were
imaged bright, contrast variations due to the exchange force are
expected because a reduction of the symmetry at surface sites
leads to a magnetic moment of the oxygen atoms, too. But this
moment is estimated to be less then 10\,\% of the one over the
nickel sites, so that the exchange effect is expected to be much
less pronounced \cite{magnetic_oxygen}.

\section{Results and analysis}

The NiO(001) surface was investigated with the improved
cantilevers, which allow stable imaging at oscillation amplitudes
as small as $1\,\rm \AA{}$ and carry three different types of tips
- nonmagnetic W tips, ferromagnetic Co tips and antiferromagnetic
NiO tips (see Fig.\,3a)). A large scale scan reveals step
structures as shown in Fig.\,2. The (001) surfaces are not ideal -
a few screw dislocations are visible - but flat terraces with a
width between 0.05 and 0.5\,$\rm \mu m$ provide a good basis for
atomic resolution.

\begin{figure}
\begin{center}
\includegraphics[clip=true,width=0.38\textwidth]{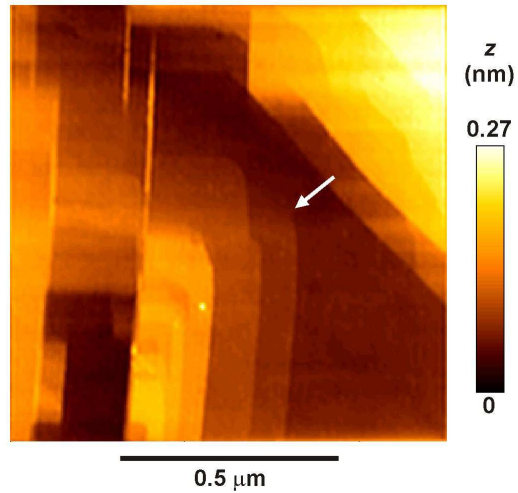}
\end{center}
\caption{(color online) Large scale step structure on NiO(001)
revealed with FM-AFM equipped with a NiO tip ($A\approx 2\,\rm
\AA{}$, $\Delta f=+15\,$Hz). Between wide flat terraces few screw
dislocations are visible, such as the one indicated by a white
arrow.}
\end{figure}

Figure 3 shows that atomic contrast on flat and clean NiO(001)
surfaces was obtained with all three kinds of tips (W, Co and NiO
tips in b), c) and d), respectively). The exact parameters of the
cantilevers are listed in table I.
\begin{table}
\begin{tabular}{|c||c|c|c|}
    \hline
                    & $f_0$ (Hz) & $k$ (N/m)
                        \\
    \hline \hline
    W tip           & 30675     & 3690              \\
    \hline
    Co tip          & 40535     & 3540              \\
    \hline
    NiO tip         & 43618     & 4020             \\
    \hline
\end{tabular}
\caption{Eigenfrequency and stiffness of the force sensors used in
the experiments.}
\end{table}

The images were acquired at $\Delta f = -20$\,Hz, $-23$\,Hz and
$-25$\,Hz with $A\approx 1\rm\,\AA{}$. Therefore, the normalized
frequency shift $\gamma = \Delta fkA^{3/2}/f_0$ was
$-2.4$\,fN$\sqrt{\rm m}$, $-2.0$\,fN$\sqrt{\rm m}$ and
$-2.3$\,fN$\sqrt{\rm m}$, respectively. Neighboring protrusions
are spaced by roughly $4\,\rm \AA{}$, indicating that only one
sort of atoms is imaged. A corrugation of around 25\,pm is
observed in these topographical images. The chemical bonding
forces responsible for the atomic resolution are assumed to be on
the order of $F_{\rm chem}\approx 1\,$nN \cite{giessibl_05} -- ten
times larger than the expected exchange force ($F_{\rm ex}\approx
0.1$\,nN, see above). Contributions of the exchange interaction to
the total tip-sample force are expected to cause about 10 percent
of the total atomic corrugation. Because it is not clear whether
Ni or O appears as a maximum and the exchange corrugation is
expected to be maximal on top of Ni, we have to analyze both,
maxima and minima, in line profiles.

\begin{figure}
\begin{center}
\includegraphics[clip=true,width=0.48\textwidth]{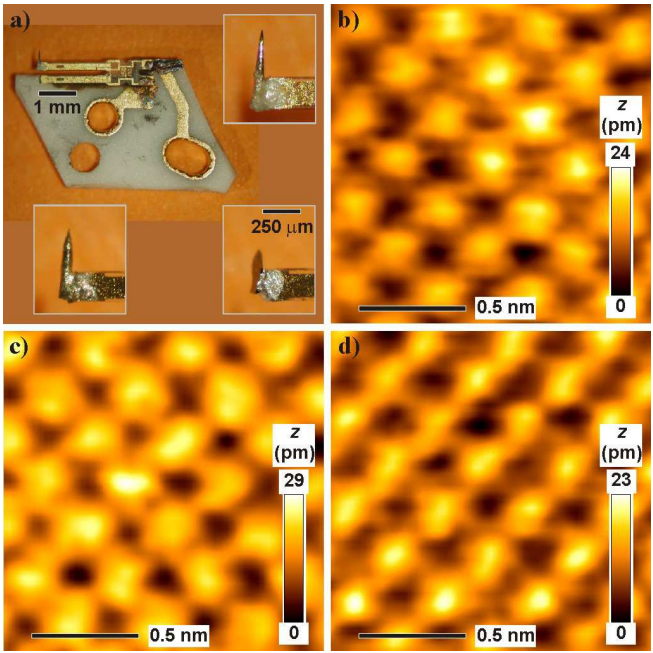}
\end{center}
\caption{(color online) qPlus sensors with tungsten (top), cobalt
(bottom left) and nickel oxide (bottom right) tips as shown in a)
allow FM-AFM with atomic resolution on NiO(001) surfaces; imaging
parameters: $A\approx 1\,\rm \AA{}$ and b) $\Delta f=-20\,$Hz (W
tip), c) $\Delta f=-23\,$Hz (Co tip), d) $\Delta f=-25\,$Hz (NiO
tip).}
\end{figure}

A detailed investigation of the correlation between the imaging
parameters and the corrugation (no images shown here) corroborates
the intuitive expectations: Decreasing the oscillation amplitude
leads to a much clearer resolution and, in addition, to an
increased corrugation. Decreasing the set point of the frequency
shift $\Delta f$ causes a further approach to the sample surface.
Hence, a greater influence of the short-range forces that lead to
the atomic resolution is expected. Indeed the corrugation in
associated height profiles of a corresponding series increases
with increasing magnitude of the frequency shift setpoint. These
measurements demonstrated that a small amplitude and a large
frequency shift are key issues for good atomic resolution.

Therefore, we continuously decreased the frequency shift ($\Delta
f<0$) while imaging at small amplitudes. Because of the large
stiffness of the modified sensors and the careful choice of the
tip material we were able to reach the repulsive regime, where
$\Delta f>0$. For the first time, atomic resolution of NiO(001)
surfaces with a positive frequency shift, i.e. operation at a
distance at or closer than the interatomic distance in bulk NiO,
was achieved. It is important to note that we used $\log{|\Delta
f|}$ as a feedback signal, but we recorded $\Delta f$ as well to
confirm the sign of $\Delta f$ (see Ref. \cite{caf_giessibl} for
more details). In Fig.\,4a) a topographical image taken with a NiO
tip at $\Delta f = +66$\,Hz and $A\approx 1\rm\,\AA{}$, i.e.\
$\gamma = +2.8\rm \,fN\sqrt{m}$, is presented. Simultaneously, the
dissipation was recorded and the result is shown in Fig.\,4b). The
damping is determined from the driving amplitude that is necessary
to keep the total energy of the cantilever constant. Variations in
the dissipation therefore correspond to changes in the energy of
the interactions \cite{garcia}. Consequently, influences of the
exchange force are expected to be detectable via the attributed
changes in energy over adjacent atom sites in the dissipation
channel, too. However, estimations yield that the ratio $E_{\rm
ex}/E_{\rm chem}$ is less favorable than the one of the forces
$F_{\rm ex}/F_{\rm chem} \approx 1/10$.

\begin{figure}
\begin{center}
\includegraphics[clip=true,width=0.48\textwidth]{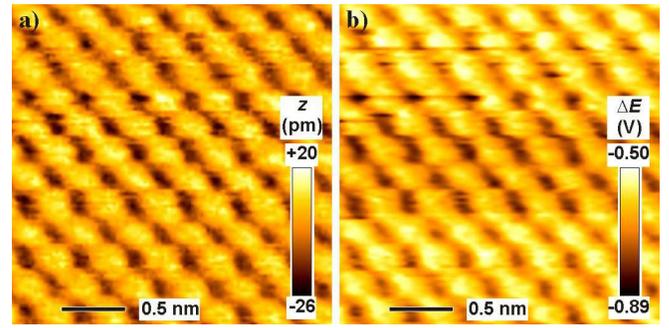}
\end{center}
\caption{(color online) Atomic resolution on NiO(001) obtained
with FM-AFM in the repulsive mode. The images were taken with a
NiO tip oscillating with $A\approx 1\,\rm \AA{}$ at $\Delta f
=+66\,$Hz. a) is a topographical picture whereas b) presents the
damping signals.}
\end{figure}

In total, a large variety of images was acquired showing atomic
resolution with different tips in various distance regimes and
several channels. As shown in Fig. 2, screw dislocations are
present on this sample. We expect that screw dislocations alter
the spin order, and even if spin alignment between tip and sample
may be weak on one region, with all the surface regions that have
been scanned there should be one region where spin alignment
between tip and sample is sufficient to observe spin contrast.
Possible spin order was searched by taking line profiles along the
two directions of the diagonals and subsequent comparison. A more
sensitive analysis method is offered by fast Fourier
transformation (FFT) of the topographical images. The expected
antiferromagnetic spin order of NiO (001) should reveal itself by
a peak at half the spatial frequency of the fundamental lattice,
thus additional Fourier peaks at ($\frac{1}{2 a_0}$,$\frac{1}{2
a_0}$) or ($\frac{1}{2 a_0}$,-$\frac{1}{2 a_0}$) should appear.
The inset in Fig.\,5 (a) presents the Fourier image of the main
topographical image that was acquired with a NiO tip at $\gamma =
-9.0\rm \,fN\sqrt{m}$ ($\Delta f = -98$\,Hz and $A\approx
1\rm\,\AA{}$). The expected additional peaks are not present in
Figure . When integrating the intensity $I$ of the FFT image over
areas A, A', B, and B', the ratio
$(I_A+I_{A'}-I_B-I_{B'})/(I_A+I_{A'}+I_B+I_{B'})$ is a measure of
spin polarization. In Figure 5 a) and the other images taken with
W or Co tips, this ratio is approximately zero. In Figure 5 b),
the spin polarization is $\approx -10$\,\%. Figure 5a) and 5b)
were taken with a NiO tip, but at a slightly different lateral
position and after a tip change that revealed itself by a glitched
line and a contrast change. The main peaks at
($\pm\frac{1}{a_0}$,$\pm\frac{1}{a_0}$)in the FFT image insets in
Fig. 5 a) and b) have a height of 17 arbitrary units (a.u.), while
the root-mean-square (rms) noise floor is at 3.0 a.u. and the
areas B and B' are at 3.5 a.u. rms. In total, the superstructure
has a corrugation of roughly
$\sqrt{3.5^2-3.0^2})/\sqrt{17^2-3.0^2}\times 25$\,pm = 2.8\,pm --
too small to be seen in the real space image but noticeable in the
FFT image. Hence, the experimental spin corrugation amplitude is
11\,\% of the fundamental currugation, somewhat less than the
16\,\% estimate presented in the experimental section.
\begin{figure}
\begin{center}
\includegraphics[clip=true,width=0.38\textwidth]{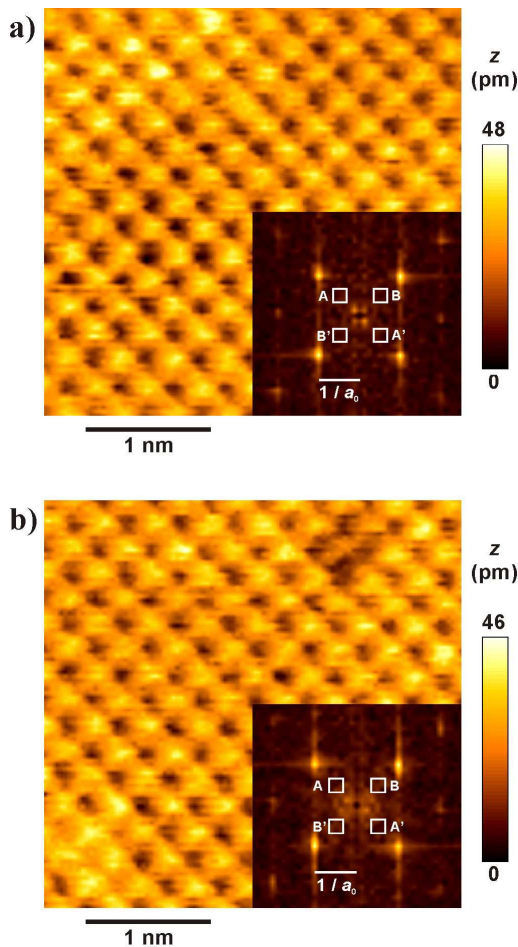}
\end{center}
\caption{(color online) a) FM-AFM image of a NiO(001) surface
taken with a NiO tip at $A\approx 1\,\rm\AA{}$ and $\Delta
f=-98\,$Hz. The presence of the two defects in the upper right and
in the lower left corner shows that true atomic resolution is
obtained, i.e. a single tip atom is responsible for imaging. The
inset shows the central section of the Fourier transform of the
topographical image. A peak at half the spatial frequency of one
of the two base peaks would be visible if the contribution of the
exchange interaction was larger than instrumental noise (see
text). Here,
$(I_A+I_{A'}-I_B-I_{B'})/(I_A+I_{A'}+I_B+I_{B'})\approx 0$. b)
Example where spin contrast appears to be present. The inset also
shows the Fourier transformed image, where
$(I_A+I_{A'}-I_B-I_{B'})/(I_A+I_{A'}+I_B+I_{B'})\approx -0.1$ (see
text). The data presented in a) and b) was taken within the same
measurement session, but a tip change indicated by a glitch and an
overall contrast change had occurred between the images.}
\end{figure}

\section{Discussion}

The small extent of apparent spin contrast in most experimental
images is puzzling. When assuming perfect spin alignment between
tip and sample, the expected contribution of the exchange
interaction is 16 times larger than the estimated instrumental
noise level and 5 times larger than the observed best-case
polarization. Calculations have shown that very small tip-sample
distances are necessary to observe spin contrast even though tip
ion instabilities may result at very small distance
\cite{Foster_2001}. Here, we have been able to image in the
repulsive regime with positive frequency shifts and generally at
distances close to the bulk neighbor distance, where optimal spin
contrast is expected \cite{Momida_2005}. While tips remained
stable, we did not observe spin contrast in the expected
magnitude. Stable imaging at a short tip-sample spacing with the
ferromagnetic Co tips was possible because of Co's moderate
reactivity with NiO. For revealing short-range magnetic forces,
another parameter is highly important in addition to the
tip-sample spacing, the relative orientation of the interacting
spins. Ideally, tip and sample spin are aligned (anti-) parallel,
but a misalignment of $60^{\circ}$ is expected to yield half the
maximal spin contrast (compare Fig. 6a)). Because there are six
possible orientations for the spins in the NiO crystal and because
we imaged large areas containing symmmetry-breaking screw
dislocations, for a given direction of the tip spin one domain has
to exist where the deviation of the relative orientation of the
spins is $60^{\circ}$ at most. Considerations of the statistical
partition of the spin alignments yield this maximum misalignment
angle, too. When imaging NiO(001) the position on the surface and
accordingly the investigated magnetic domain was changed multiple
times. Therefore, we assume to find adequate spin alignment in
several cases - at least for a limited time as we can not rule out
spin flips of the tip but also within the sample during the scan.

\begin{figure}
\begin{center}
\includegraphics[clip=true,width=0.48\textwidth]{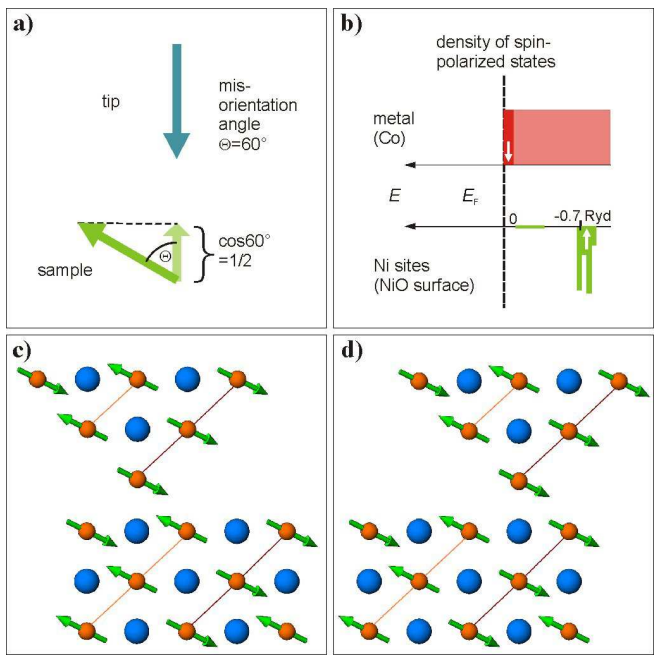}
\end{center}
\caption{(color online) a) Sketch illustrating the effect of a
misalignment between tip and sample spin; b) comparison of density
of spin-polarized states of the NiO surface and a metal tip
revealing the improbability of an interaction; c) sketch of a Ni
front atom located above a O sample atom such that spin order is
preserved from tip to sample d) a Ni front atom also sits on top
of a O sample atom, but the spin order is broken. }
\end{figure}

A last consideration regards the tip material. The expected spin
contrast originates in the exchange interaction, which is not due
to a magnetic dipole-dipole interaction, but due to
spin-controlled electrostatic interaction. Exchange interaction
can only happen if a bonding orbital between tip and sample
evolves, i.e. if an electronic state at a given energy has a large
probability amplitude in both tip and sample atoms. The energy of
the spin-polarized states located at the Ni sites of the NiO
surface is about 0.7\,Ryd below the Fermi level
\cite{NiO_exchange_interactions}. Approaching a metal tip \mbox{-
for example made from cobalt -} to the NiO surface the Fermi
levels will match. The spin-polarized states of the metal form a
small band under the Fermi level with a bandwidth much smaller
than the estimated 10\,eV energy difference to the NiO surface.
Consequently the formation of a molecular orbital that has a large
amplitude on both tip and NiO sample appears to be unlikely
(compare Fig.\,6b)). The result of this discussion is that
ferromagnetic metal tips may not be the optimal choice for
detecting exchange forces with a NiO sample. In spin-polarized
tunneling, the physical mechanism behind spin contrast is
different: the tunneling current is proportional to the
spin-dependent density of states in tip and sample, and because
electrons are tunneling from states close to the Fermi level in
the tip (sample) into states close to the Fermi level in the
sample (tip), the energetic equality is automatically fulfilled.

In order to obtain optimal tip-sample interactions we have chosen
an approach that is conceptually very simple: we manufacture a tip
of NiO and measure its interaction with a NiO surface. If the spin
orientation in tip and sample was parallel, the spin orientation
depicted in Fig. 6 c), where the spin order continues from tip to
sample, is energetically lower than the one shown in Fig. 6 d),
where the spin order is broken. It follows that the force in case
6 c) is larger than in 6 d). Spin contrast is expected to become
visible no matter whether the tip atom is Ni or O, because the
spin order could be continued in sequence in either case and spins
between Ni ions couple through superexchange in NiO
\cite{Anderson_1950}.

\section{Summary and Outlook}
In summary, we find evidence for spin contrast on NiO (001) in
Fourier space images when using tips made from NiO and maximizing
the sensitivity to short range forces by adapting a small
amplitude scheme. Spin contrast was not detectable when using
magnetized Co tips, and we have provided a qualitative explanation
by arguing that the energy of the spin polarized states in Co tips
and Ni ions on NiO does not match. This argument may explain why
clear spin contrast has not been observed in very low noise
experiments of other groups conducted at low temperatures. It is
expected that the antiferromagnetic spin order of NiO (001) is
fully developed at room temperature
\cite{NiO_exchange_interactions}, but the tip atom of a sharp NiO
crystallite may require much lower temperatures to develop spin
order than the bulk. Therefore, we anticipate that the spin
contrast signal will become stronger at low temperatures.
Repeating this experiment at low temperatures with NiO tips will
result in lower noise, such that spin order not only shows in
Fourier images, but in real space images as well. Also, tip
preparation can be improved. It would be beneficial to cleave the
tips in situ in ultrahigh vacuum just as the sample. In this case,
the tip would be definitely clean and uncontaminated, at least at
the beginning of the experiment. The use of amplitudes in the
\AA{}- and sub-\AA{}- regime was only possible by building force
sensors with a stiffness on the order of 4\,kN/m. Due to a careful
choice of tip material, a small tip-sample distance could be
realized without losing atomic contrast. These are essential
requirements for detecting the extreme short-range exchange
interaction between a magnetic tip and the antiferromagnetic
sample surface. For the future, we plan to perform these
measurements at low temperatures and to utilize advanced tip
preparation methods such as {\it in situ} tip cleaving. The use of
tips made from the same material as the sample has proven to be
very successful. This concept may be transferable to other
systems, expanding the conceptual beauty of break-junction
experiments\cite{Agrait_2003} to three dimensional imaging.

\section{Acknowledgments}
We thank T. Eguchi, Ch. Schiller and M. Breitschaft for helpful
discussions and support. This work is supported by the
Bundesministerium f\"{u}r Bildung und Forschung (project
EKM13N6918) and by the European Science Foundation (THIOX).

\bibliography{lit}

\begin{thebibliography}{31}
\expandafter\ifx\csname natexlab\endcsname\relax\def\natexlab#1{#1}\fi
\expandafter\ifx\csname bibnamefont\endcsname\relax
  \def\bibnamefont#1{#1}\fi
\expandafter\ifx\csname bibfnamefont\endcsname\relax
  \def\bibfnamefont#1{#1}\fi
\expandafter\ifx\csname citenamefont\endcsname\relax
  \def\citenamefont#1{#1}\fi
\expandafter\ifx\csname url\endcsname\relax
  \def\url#1{\texttt{#1}}\fi
\expandafter\ifx\csname urlprefix\endcsname\relax\def\urlprefix{URL }\fi
\providecommand{\bibinfo}[2]{#2}
\providecommand{\eprint}[2][]{\url{#2}}

\bibitem[{\citenamefont{{G. Baym}}(1969)}]{Baym}
\bibinfo{author}{\bibnamefont{{G. Baym}}}, \emph{\bibinfo{title}{Lectures on
  Quantum Mechanics}} (\bibinfo{publisher}{W. A. Benjamin, Inc.},
  \bibinfo{address}{New York}, \bibinfo{year}{1969}).

\bibitem[{\citenamefont{Bode}(2003)}]{Bode_2003}
\bibinfo{author}{\bibfnamefont{M.}~\bibnamefont{Bode}}, \bibinfo{journal}{Rep.
  Prog. Phys.} \textbf{\bibinfo{volume}{66}}, \bibinfo{pages}{523}
  (\bibinfo{year}{2003}).

\bibitem[{\citenamefont{Heinze et~al.}(2000)\citenamefont{Heinze, Bode,
  Kubetzka, Pietzsch, Nie, Bl\ügel, and Wiesendanger}}]{Heinze_2000}
\bibinfo{author}{\bibfnamefont{S.}~\bibnamefont{Heinze}},
  \bibinfo{author}{\bibfnamefont{M.}~\bibnamefont{Bode}},
  \bibinfo{author}{\bibfnamefont{A.}~\bibnamefont{Kubetzka}},
  \bibinfo{author}{\bibfnamefont{O.}~\bibnamefont{Pietzsch}},
  \bibinfo{author}{\bibfnamefont{X.}~\bibnamefont{Nie}},
  \bibinfo{author}{\bibfnamefont{S.}~\bibnamefont{Bl\ügel}}, \bibnamefont{and}
  \bibinfo{author}{\bibfnamefont{R.}~\bibnamefont{Wiesendanger}},
  \bibinfo{journal}{Science} \textbf{\bibinfo{volume}{288}},
  \bibinfo{pages}{1805} (\bibinfo{year}{2000}).

\bibitem[{\citenamefont{Heinrich et~al.}(2004)\citenamefont{Heinrich, Gupta,
  Lutz, and Eigler}}]{Heinrich_2004}
\bibinfo{author}{\bibfnamefont{A.~J.} \bibnamefont{Heinrich}},
  \bibinfo{author}{\bibfnamefont{J.~A.} \bibnamefont{Gupta}},
  \bibinfo{author}{\bibfnamefont{C.~P.} \bibnamefont{Lutz}}, \bibnamefont{and}
  \bibinfo{author}{\bibfnamefont{D.~M.} \bibnamefont{Eigler}},
  \bibinfo{journal}{Science} \textbf{\bibinfo{volume}{306}},
  \bibinfo{pages}{466} (\bibinfo{year}{2004}).

\bibitem[{\citenamefont{{G. Binnig} et~al.}(1986)\citenamefont{{G. Binnig},
  {C.\,F. Quate}, and {Ch. Gerber}}}]{binnig_86}
\bibinfo{author}{\bibnamefont{{G. Binnig}}},
  \bibinfo{author}{\bibnamefont{{C.\,F. Quate}}}, \bibnamefont{and}
  \bibinfo{author}{\bibnamefont{{Ch. Gerber}}}, \bibinfo{journal}{Phys. Rev.
  Lett.} \textbf{\bibinfo{volume}{56}}, \bibinfo{pages}{930}
  (\bibinfo{year}{1986}).

\bibitem[{\citenamefont{{K. Nakamura} et~al.}(1999)\citenamefont{{K. Nakamura},
  {T. Oguchi}, {H. Hasegawa}, {K. Sueoka}, {K. Hayakawa}, and {K.
  Mukasa}}}]{Nakamura_1999}
\bibinfo{author}{\bibnamefont{{K. Nakamura}}},
  \bibinfo{author}{\bibnamefont{{T. Oguchi}}},
  \bibinfo{author}{\bibnamefont{{H. Hasegawa}}},
  \bibinfo{author}{\bibnamefont{{K. Sueoka}}},
  \bibinfo{author}{\bibnamefont{{K. Hayakawa}}}, \bibnamefont{and}
  \bibinfo{author}{\bibnamefont{{K. Mukasa}}}, \bibinfo{journal}{Appl. Surf.
  Sci.} \textbf{\bibinfo{volume}{140}}, \bibinfo{pages}{366}
  (\bibinfo{year}{1999}).

\bibitem[{\citenamefont{Imada et~al.}(1309)\citenamefont{Imada, Fujimori, and
  Tokura}}]{Metal_Insulator_Transitions_1998}
\bibinfo{author}{\bibfnamefont{M.}~\bibnamefont{Imada}},
  \bibinfo{author}{\bibfnamefont{A.}~\bibnamefont{Fujimori}}, \bibnamefont{and}
  \bibinfo{author}{\bibfnamefont{Y.}~\bibnamefont{Tokura}},
  \bibinfo{journal}{Rev. Mod. Phys.} \textbf{\bibinfo{volume}{70}},
  \bibinfo{pages}{4} (\bibinfo{year}{1309}).

\bibitem[{\citenamefont{{H. Hosoi} et~al.}(2000)\citenamefont{{H. Hosoi}, {K.
  Sueoka}, {K. Hayakawa}, and {K. Mukasa}}}]{NiO_Hosoi_2000}
\bibinfo{author}{\bibnamefont{{H. Hosoi}}}, \bibinfo{author}{\bibnamefont{{K.
  Sueoka}}}, \bibinfo{author}{\bibnamefont{{K. Hayakawa}}}, \bibnamefont{and}
  \bibinfo{author}{\bibnamefont{{K. Mukasa}}}, \bibinfo{journal}{Appl. Surf.
  Sci.} \textbf{\bibinfo{volume}{157}}, \bibinfo{pages}{218}
  (\bibinfo{year}{2000}).

\bibitem[{\citenamefont{{H. Hosoi} et~al.}(2001)\citenamefont{{H. Hosoi}, {M.
  Kimura}, {K. Hayakawa}, {K. Sueoka}, and {K. Mukasa}}}]{NiO_Hosoi_2001}
\bibinfo{author}{\bibnamefont{{H. Hosoi}}}, \bibinfo{author}{\bibnamefont{{M.
  Kimura}}}, \bibinfo{author}{\bibnamefont{{K. Hayakawa}}},
  \bibinfo{author}{\bibnamefont{{K. Sueoka}}}, \bibnamefont{and}
  \bibinfo{author}{\bibnamefont{{K. Mukasa}}}, \bibinfo{journal}{Appl. Phys. A}
  \textbf{\bibinfo{volume}{72}}, \bibinfo{pages}{S23} (\bibinfo{year}{2001}).

\bibitem[{\citenamefont{{H. Hosoi} et~al.}(2004)\citenamefont{{H. Hosoi}, {K.
  Sueoka}, and {K. Mukasa}}}]{NiO_Hosoi_2004}
\bibinfo{author}{\bibnamefont{{H. Hosoi}}}, \bibinfo{author}{\bibnamefont{{K.
  Sueoka}}}, \bibnamefont{and} \bibinfo{author}{\bibnamefont{{K. Mukasa}}},
  \bibinfo{journal}{Nanotechnology} \textbf{\bibinfo{volume}{15}},
  \bibinfo{pages}{505} (\bibinfo{year}{2004}).

\bibitem[{\citenamefont{{W. Allers} et~al.}(2001)\citenamefont{{W. Allers}, {S.
  Langkat}, and {R. Wiesendanger}}}]{NiO_Wiesendanger_2001}
\bibinfo{author}{\bibnamefont{{W. Allers}}}, \bibinfo{author}{\bibnamefont{{S.
  Langkat}}}, \bibnamefont{and} \bibinfo{author}{\bibnamefont{{R.
  Wiesendanger}}}, \bibinfo{journal}{Appl. Phys. A}
  \textbf{\bibinfo{volume}{72}}, \bibinfo{pages}{S27} (\bibinfo{year}{2001}).

\bibitem[{\citenamefont{{H. H\"olscher} et~al.}(2002)\citenamefont{{H.
  H\"olscher}, {S.\,M. Langkat}, {A. Schwarz}, and {R.
  Wiesendanger}}}]{NiO_Wiesendanger_2002}
\bibinfo{author}{\bibnamefont{{H. H\"olscher}}},
  \bibinfo{author}{\bibnamefont{{S.\,M. Langkat}}},
  \bibinfo{author}{\bibnamefont{{A. Schwarz}}}, \bibnamefont{and}
  \bibinfo{author}{\bibnamefont{{R. Wiesendanger}}}, \bibinfo{journal}{Appl.
  Phys. Lett.} \textbf{\bibinfo{volume}{81}}, \bibinfo{pages}{4428}
  (\bibinfo{year}{2002}).

\bibitem[{\citenamefont{{S.\,M. Langkat} et~al.}(2003)\citenamefont{{S.\,M.
  Langkat}, {H. H\"olscher}, {A. Schwarz}, and {R.
  Wiesendanger}}}]{NiO_Wiesendanger_2003}
\bibinfo{author}{\bibnamefont{{S.\,M. Langkat}}},
  \bibinfo{author}{\bibnamefont{{H. H\"olscher}}},
  \bibinfo{author}{\bibnamefont{{A. Schwarz}}}, \bibnamefont{and}
  \bibinfo{author}{\bibnamefont{{R. Wiesendanger}}}, \bibinfo{journal}{Surf.
  Sci.} \textbf{\bibinfo{volume}{527}}, \bibinfo{pages}{12}
  (\bibinfo{year}{2003}).

\bibitem[{\citenamefont{{R. Hoffmann} et~al.}(2003)\citenamefont{{R. Hoffmann},
  {M.\,A. Lantz}, {H.\,J. Hug}, {P.\,J.\,A. van Schendel}, {P. Kappenberger},
  {S. Martin}, {A. Baratoff}, and {H.-J. G\üntherodt}}}]{NiO_Hoffmann}
\bibinfo{author}{\bibnamefont{{R. Hoffmann}}},
  \bibinfo{author}{\bibnamefont{{M.\,A. Lantz}}},
  \bibinfo{author}{\bibnamefont{{H.\,J. Hug}}},
  \bibinfo{author}{\bibnamefont{{P.\,J.\,A. van Schendel}}},
  \bibinfo{author}{\bibnamefont{{P. Kappenberger}}},
  \bibinfo{author}{\bibnamefont{{S. Martin}}},
  \bibinfo{author}{\bibnamefont{{A. Baratoff}}}, \bibnamefont{and}
  \bibinfo{author}{\bibnamefont{{H.-J. G\üntherodt}}}, \bibinfo{journal}{Phys.
  Rev. B} \textbf{\bibinfo{volume}{67}}, \bibinfo{pages}{085402}
  (\bibinfo{year}{2003}).

\bibitem[{\citenamefont{{A.\,S. Foster} and {A.\,L.
  Shluger}}(2001)}]{NiO_expectations_Foster}
\bibinfo{author}{\bibnamefont{{A.\,S. Foster}}} \bibnamefont{and}
  \bibinfo{author}{\bibnamefont{{A.\,L. Shluger}}}, \bibinfo{journal}{Surf.
  Sci.} \textbf{\bibinfo{volume}{490}}, \bibinfo{pages}{211}
  (\bibinfo{year}{2001}).

\bibitem[{\citenamefont{{F.\,U. Hillebrecht} et~al.}(2001)\citenamefont{{F.\,U.
  Hillebrecht}, {H. Ohldag}, {N.\,B. Weber}, {C. Bethke}, {U. Mick}, {M.
  Weiss}, and {J. Bahrdt}}}]{NiO_spindirection}
\bibinfo{author}{\bibnamefont{{F.\,U. Hillebrecht}}},
  \bibinfo{author}{\bibnamefont{{H. Ohldag}}},
  \bibinfo{author}{\bibnamefont{{N.\,B. Weber}}},
  \bibinfo{author}{\bibnamefont{{C. Bethke}}},
  \bibinfo{author}{\bibnamefont{{U. Mick}}}, \bibinfo{author}{\bibnamefont{{M.
  Weiss}}}, \bibnamefont{and} \bibinfo{author}{\bibnamefont{{J. Bahrdt}}},
  \bibinfo{journal}{Phys. Rev. Lett.} \textbf{\bibinfo{volume}{86}},
  \bibinfo{pages}{3419} (\bibinfo{year}{2001}).

\bibitem[{\citenamefont{{M. Schmid} et~al.}(2006)\citenamefont{{M. Schmid}, {A.
  Renner}, and {F.\,J. Giessibl}}}]{cleaver}
\bibinfo{author}{\bibnamefont{{M. Schmid}}}, \bibinfo{author}{\bibnamefont{{A.
  Renner}}}, \bibnamefont{and} \bibinfo{author}{\bibnamefont{{F.\,J.
  Giessibl}}}, \bibinfo{journal}{Rev. Sci. Instrum.}
  \textbf{\bibinfo{volume}{77}}, \bibinfo{pages}{036101}
  (\bibinfo{year}{2006}).

\bibitem[{\citenamefont{{K. Nakamura} et~al.}(1997)\citenamefont{{K. Nakamura},
  {H. Hasegawa}, {T. Oguchi}, {K. Sueoka}, {K. Hayakawa}, and {K.
  Mukasa}}}]{NiO_expectations_Mukasa97}
\bibinfo{author}{\bibnamefont{{K. Nakamura}}},
  \bibinfo{author}{\bibnamefont{{H. Hasegawa}}},
  \bibinfo{author}{\bibnamefont{{T. Oguchi}}},
  \bibinfo{author}{\bibnamefont{{K. Sueoka}}},
  \bibinfo{author}{\bibnamefont{{K. Hayakawa}}}, \bibnamefont{and}
  \bibinfo{author}{\bibnamefont{{K. Mukasa}}}, \bibinfo{journal}{Phys. Rev. B}
  \textbf{\bibinfo{volume}{56}}, \bibinfo{pages}{3218} (\bibinfo{year}{1997}).

\bibitem[{\citenamefont{{D. K\"odderitzsch} et~al.}(2002)\citenamefont{{D.
  K\"odderitzsch}, {W. Hergert}, {W.\,M. Temmerman}, {Z. Szotek}, {A. Ernst},
  and {H. Winter}}}]{NiO_exchange_interactions}
\bibinfo{author}{\bibnamefont{{D. K\"odderitzsch}}},
  \bibinfo{author}{\bibnamefont{{W. Hergert}}},
  \bibinfo{author}{\bibnamefont{{W.\,M. Temmerman}}},
  \bibinfo{author}{\bibnamefont{{Z. Szotek}}},
  \bibinfo{author}{\bibnamefont{{A. Ernst}}}, \bibnamefont{and}
  \bibinfo{author}{\bibnamefont{{H. Winter}}}, \bibinfo{journal}{Phys. Rev. B}
  \textbf{\bibinfo{volume}{66}}, \bibinfo{pages}{064434}
  (\bibinfo{year}{2002}).

\bibitem[{\citenamefont{Lide}(1996)}]{CRC}
\bibinfo{editor}{\bibfnamefont{D.~R.} \bibnamefont{Lide}}, ed.,
  \emph{\bibinfo{title}{CRC Handbook of Chemistry and Physics}}
  (\bibinfo{publisher}{CRC Press}, \bibinfo{address}{Boca Raton},
  \bibinfo{year}{1996}), \bibinfo{edition}{77th} ed.

\bibitem[{\citenamefont{{F.\,J. Giessibl}}(2003{\natexlab{a}})}]{SNR}
\bibinfo{author}{\bibnamefont{{F.\,J. Giessibl}}}, \bibinfo{journal}{AIP
  Conference Proceedings} \textbf{\bibinfo{volume}{696}}, \bibinfo{pages}{60}
  (\bibinfo{year}{2003}{\natexlab{a}}).

\bibitem[{\citenamefont{{F.\,J. Giessibl}}(2003{\natexlab{b}})}]{basic}
\bibinfo{author}{\bibnamefont{{F.\,J. Giessibl}}}, \bibinfo{journal}{Rev. Mod.
  Phys.} \textbf{\bibinfo{volume}{75}}, \bibinfo{pages}{949}
  (\bibinfo{year}{2003}{\natexlab{b}}).

\bibitem[{\citenamefont{{F.\,J. Giessibl} and {H.
  Bielefeldt}}(2000)}]{giessibl_00}
\bibinfo{author}{\bibnamefont{{F.\,J. Giessibl}}} \bibnamefont{and}
  \bibinfo{author}{\bibnamefont{{H. Bielefeldt}}}, \bibinfo{journal}{Phys. Rev.
  B} \textbf{\bibinfo{volume}{61}}, \bibinfo{pages}{9968}
  (\bibinfo{year}{2000}).

\bibitem[{\citenamefont{Momida and Oguchi}(2005)}]{Momida_2005}
\bibinfo{author}{\bibfnamefont{H.}~\bibnamefont{Momida}} \bibnamefont{and}
  \bibinfo{author}{\bibfnamefont{T.}~\bibnamefont{Oguchi}},
  \bibinfo{journal}{Surf. Sci.} \textbf{\bibinfo{volume}{590}},
  \bibinfo{pages}{42} (\bibinfo{year}{2005}).

\bibitem[{\citenamefont{{H. Momida} and {T. Oguchi}}(2003)}]{magnetic_oxygen}
\bibinfo{author}{\bibnamefont{{H. Momida}}} \bibnamefont{and}
  \bibinfo{author}{\bibnamefont{{T. Oguchi}}}, \bibinfo{journal}{J. Phys. Soc.
  Jpn.} \textbf{\bibinfo{volume}{72}}, \bibinfo{pages}{588}
  (\bibinfo{year}{2003}).

\bibitem[{\citenamefont{{F.\,J. Giessibl}}(2005)}]{giessibl_05}
\bibinfo{author}{\bibnamefont{{F.\,J. Giessibl}}}, \bibinfo{journal}{Materials
  Today} \textbf{\bibinfo{volume}{8(5)}}, \bibinfo{pages}{32}
  (\bibinfo{year}{2005}).

\bibitem[{\citenamefont{{F.\,J. Giessibl} and {M.
  Reichling}}(2005)}]{caf_giessibl}
\bibinfo{author}{\bibnamefont{{F.\,J. Giessibl}}} \bibnamefont{and}
  \bibinfo{author}{\bibnamefont{{M. Reichling}}},
  \bibinfo{journal}{Nanotechnology} \textbf{\bibinfo{volume}{16}},
  \bibinfo{pages}{S118} (\bibinfo{year}{2005}).

\bibitem[{\citenamefont{{R. Garc\'{i}a} and {R. P\'{e}rez}}(2002)}]{garcia}
\bibinfo{author}{\bibnamefont{{R. Garc\'{i}a}}} \bibnamefont{and}
  \bibinfo{author}{\bibnamefont{{R. P\'{e}rez}}}, \bibinfo{journal}{Surf. Sci.
  Rep.} \textbf{\bibinfo{volume}{47}}, \bibinfo{pages}{197}
  (\bibinfo{year}{2002}).

\bibitem[{\citenamefont{Foster and Shluger}(2001)}]{Foster_2001}
\bibinfo{author}{\bibfnamefont{A.~S.} \bibnamefont{Foster}} \bibnamefont{and}
  \bibinfo{author}{\bibfnamefont{A.~L.} \bibnamefont{Shluger}},
  \bibinfo{journal}{Surf. Sci.} \textbf{\bibinfo{volume}{490}},
  \bibinfo{pages}{211} (\bibinfo{year}{2001}).

\bibitem[{\citenamefont{Anderson}(1950)}]{Anderson_1950}
\bibinfo{author}{\bibfnamefont{P.~W.} \bibnamefont{Anderson}},
  \bibinfo{journal}{Phys. Rev.} \textbf{\bibinfo{volume}{79}},
  \bibinfo{pages}{350} (\bibinfo{year}{1950}).

\bibitem[{\citenamefont{Agrait et~al.}(2003)\citenamefont{Agrait, Yeyati, and
  van Ruitenbeek}}]{Agrait_2003}
\bibinfo{author}{\bibfnamefont{N.}~\bibnamefont{Agrait}},
  \bibinfo{author}{\bibfnamefont{A.}~\bibnamefont{Yeyati}}, \bibnamefont{and}
  \bibinfo{author}{\bibfnamefont{J.}~\bibnamefont{van Ruitenbeek}},
  \bibinfo{journal}{Phys. Rep.} \textbf{\bibinfo{volume}{377}},
  \bibinfo{pages}{81} (\bibinfo{year}{2003}).

\end{thebibliography}

\end{document}